\documentclass[preprint]{aastex}
\usepackage{color}
\usepackage{graphicx}

\slugcomment{Revised on 2014 May 12 07:30PST}

\shorttitle{Physical Properties of Asteroids in Comet--Like Orbits\\
in Infrared Asteroidal Survey Catalogs}
\shortauthors{Kim et al.} 

\makeatletter
\AtBeginDocument{\@ifpackageloaded{natbib}{\ifNAT@numbers\if@filesw\immediate\write\@auxout{\string\global\string\NAT@numberstrue}\fi\fi}{}}
\makeatother
\usepackage{amsmath}

\begin{document}

\title{Physical Properties of Asteroids in Comet-like Orbits\\
in Infrared Asteroid Survey Catalogs}

\author{Yoonyoung \textsc{Kim}, Masateru \textsc{Ishiguro}\altaffilmark{1}}
\affil{Department of Physics and Astronomy, Seoul National University,
Gwanak, Seoul 151-742, South Korea}

\and

\author{Fumihiko \textsc{Usui}}
\affil{Department of Astronomy, Graduate School of Science,  University of Tokyo,\\
7-3-1 Hongo, Bunkyo-ku, Tokyo 113-0033, Japan}

\altaffiltext{1}{Visiting Scientist, Department of Earth, Planetary and Space
Sciences, University of California at Los Angeles, 595 Charles Young
Drive East, Los Angeles, CA 90095-1567, USA}

\begin{abstract}
We investigated the population of asteroids in comet-like orbits
using available asteroid size and albedo catalogs of data taken with 
the \emph{Infrared Astronomical Satellite}, \emph{AKARI}, and the \emph{Wide-field Infrared Survey Explorer} 
on the basis of their orbital properties (i.e., the Tisserand parameter with respect to Jupiter, 
$T_\mathrm{J}$, and the aphelion distance, $Q$).
We found that (i) there are 123 asteroids in comet-like orbits by our criteria (i.e., $Q<4.5$~AU and $T_\mathrm{J}<3$),
(ii) 80\%\ of them have low albedo, $p_\mathrm{v}<0.1$, consistent with comet nuclei, (iii) low-albedo objects
among them have a size distribution shallower than that of active comet nuclei, that is,
the power index of the cumulative size distribution of around 1.1, (iv) unexpectedly, a considerable number
(i.e., 25 by our criteria) of asteroids in comet-like orbits have high albedo, $p_\mathrm{v}>0.1$. We noticed that such
high-albedo objects mostly consist of small ($D< 3$~km) bodies distributed in near-Earth space (with perihelion
distance of $q<1.3$~AU). We suggest that such high-albedo, small objects were susceptible to the
Yarkovsky effect and drifted into comet-like orbits via chaotic resonances with planets.
\end{abstract}


% SECTION 1
\section{Introduction}

% motivation (introduction)
Comets, which consist of volatiles and dark (of typical geometric albedo $p_\mathrm{v}$= 0.02--0.06; 
\citealt{Campins2000,Lamy2004}), reddish refractory, lose their volatiles near their
surfaces after many returning orbits. Simultaneously, their surfaces are covered with
an inert dust mantle that  prevents sublimation of subsurface ice 
\citep{Prialnik1988,Rickman1990}. Eventually, their appearances would be
indistinguishable from asteroids through astronomical observations. It has been speculated that
there could be dormant or extinct comets in the list of known
asteroids. Because the physical lifetime of short-period comets in the
inner Solar System ($\approx$1$\times$10$^{4}$ years) is 10--1000 times
longer than that of the devolatilization timescale of ices
\citep{Levison1997}, it is not surprising that there would be hidden comets in the
list of asteroids \citep{Weissman2002,Jewitt2004}. 

% previous researches (development)
Identification of such dormant comets from telescopic observations is
not simple because comet nuclei have a wide range of optical properties (i.e., albedos and reflectance spectra)
that overlap those of some classes of asteroids. 
Meanwhile, the Tisserand parameter, $T_\mathrm{J}$, derived from the
Jacobi integral of the circular, restricted three-body problem of the Sun, Jupiter and an interplanetary body,
provides a useful criterion for distinguishing  comets from asteroids. It is defined by
\begin{equation}
T_\mathrm{J} = \frac{a_\mathrm{J}}{a} + 2 \left[(1 -
                                e^2)\frac{a}{a_\mathrm{J}}\right]^{1/2} \cos i, 
\end{equation}
\noindent where $a_\mathrm{J}$ ($=5.2$~AU) is the semimajor axes of Jupiter, and
$a$, $e$, and $i$ are the semimajor axis, eccentricity, and inclination, respectively,
of an interplanetary body such as a comet or asteroid. 
Comets usually have $T_\mathrm{J}<3$, whereas most asteroids have
$T_\mathrm{J}>3$ \citep{Levison1997}.

An early survey of dormant comets was performed by 
\citet{Fernandez2001,Fernandez2005}. They conducted optical and infrared
observations from ground to derive the albedo of asteroids in cometary
orbits and found a clear correlation between the Tisserand parameter and
albedo, suggesting that $T_\mathrm{J}$ is a good indicator to discriminate 
between asteroids and comets.
A follow-up survey was performed by \citet{Licandro2008}, using
visible (0.55--0.90~\micron) and near-infrared (0.8--2.3~\micron)
spectrographs. They determined the spectral taxonomic types of 24
asteroids in comet-like orbits and found that all observed objects with
$T_\mathrm{J}<2.9$ have neutral or reddish spectra compatible with
comet nuclei. In addition, \citet{DeMeo2008} acquired spectra of 20 near-Earth
asteroids (defined by those having a perihelion distance $q<1.3$~AU) in comet-like
orbits. They estimated that $\sim$8\% of the observed 
near-Earth objects have surface spectral types consistent with comets. 

% contents of the paper (twist+conclusion)
Data taken with infrared space telescopes open up a
further possibility to study dormant comets. At present, three
infrared asteroid catalogs of data taken with infrared space surveyors are
available, providing information about sizes and albedos that are useful
in diagnosing the physical properties of dormant comets as well as asteroids.
The principal aim of this study is to investigate the population of
asteroids in comet-like orbits with these infrared catalogs, following the
research of \citet{Fernandez2001,Fernandez2005}.
In this paper, we adopt the term  ``asteroid in comet-like orbit" (ACO) as one having 
$T_\mathrm{J}<3.0$ and an aphelion distance $Q>4.5$~AU. The term has been occasionally used in
some  papers \citep{Fernandez2001,Fernandez2005,Licandro2006,Licandro2008}.
Moreover, we introduce the term  ``potential dormant comet" (PDC) as one
having low albedo $(p_\mathrm{v} < 0.1)$ among ACOs. The second term is a paronomasia
associating the spectra of potential dormant comets  with similar spectra of
P-type, D-type, or C-type asteroids \citep{Licandro2008,DeMeo2008}.
In Section 2, we describe the data sets and our extraction method for ACOs.
In Section 3, we show our results on the physical properties of ACOs and PDCs.
Finally, we discuss our findings and compare them with previous research in Section 4. 

% SECTION 2
\section{Applied Data and Methodology}
\subsection{Infrared Asteroid Catalogs}

We used infrared asteroid databases compiled from three infrared
all-sky surveyors, the \emph{Infrared Astronomical Satellite} (\emph{IRAS}; \citealt{Neugebauer1984}), \emph{AKARI}
\citep{Murakami2007}, and the \emph{Wide-field Infrared Survey Explorer} (\emph{WISE}; \citealt{Wright2010}).
Detailed descriptions of the asteroid catalogs compiled from these surveyors can be found in \cite{Tedesco2002}, \cite{Usui2011}, and \cite{Mainzer2011a}, respectively, 
and their series of papers.
These catalog data are available
online.\footnote{http://sbn.psi.edu/pds/resource/imps.html} \footnote{http://darts.jaxa.jp/ir/AKARI/catalogue/AcuA.html}
%In total, we obtained sizes and albedos of 2,470 asteroids (\emph{IRAS}), 5,120 asteroids (\emph{AKARI}), and 137,837 asteroids (\emph{WISE}).

\citet{Usui2014} compared these three infrared asteroidal catalogs with valid sizes and albedos
and  merged them into single catalog (I--A--W). They archived 138,285 asteroids
with sizes and albedos, detected with either \emph{IRAS}, \emph{AKARI}, or \emph{WISE} in I--A--W.
A number of asteroids were detected by two or three satellites: 1993 asteroids
by three satellites, 2812 asteroids by \emph{AKARI} and \emph{WISE}, and 312 asteroids by \emph{IRAS} and \emph{WISE}. In such cases, \citet{Usui2014} selected data from \emph{AKARI} as the highest priority, \emph{WISE}
as the second, and \emph{IRAS} as the third priority, although there are no remarkable differences in sizes and
albedos among these catalogs.
\emph{AKARI} data were given highest priority because its data have less uncertainty  than \emph{WISE} in sizes and albedos  for the largest asteroids. 
We applied the I--A--W catalog for the analysis of ACOs.
\subsection{Data Processing}

\begin{figure}
\begin{center} 
    \epsscale{0.60}
   \plotone{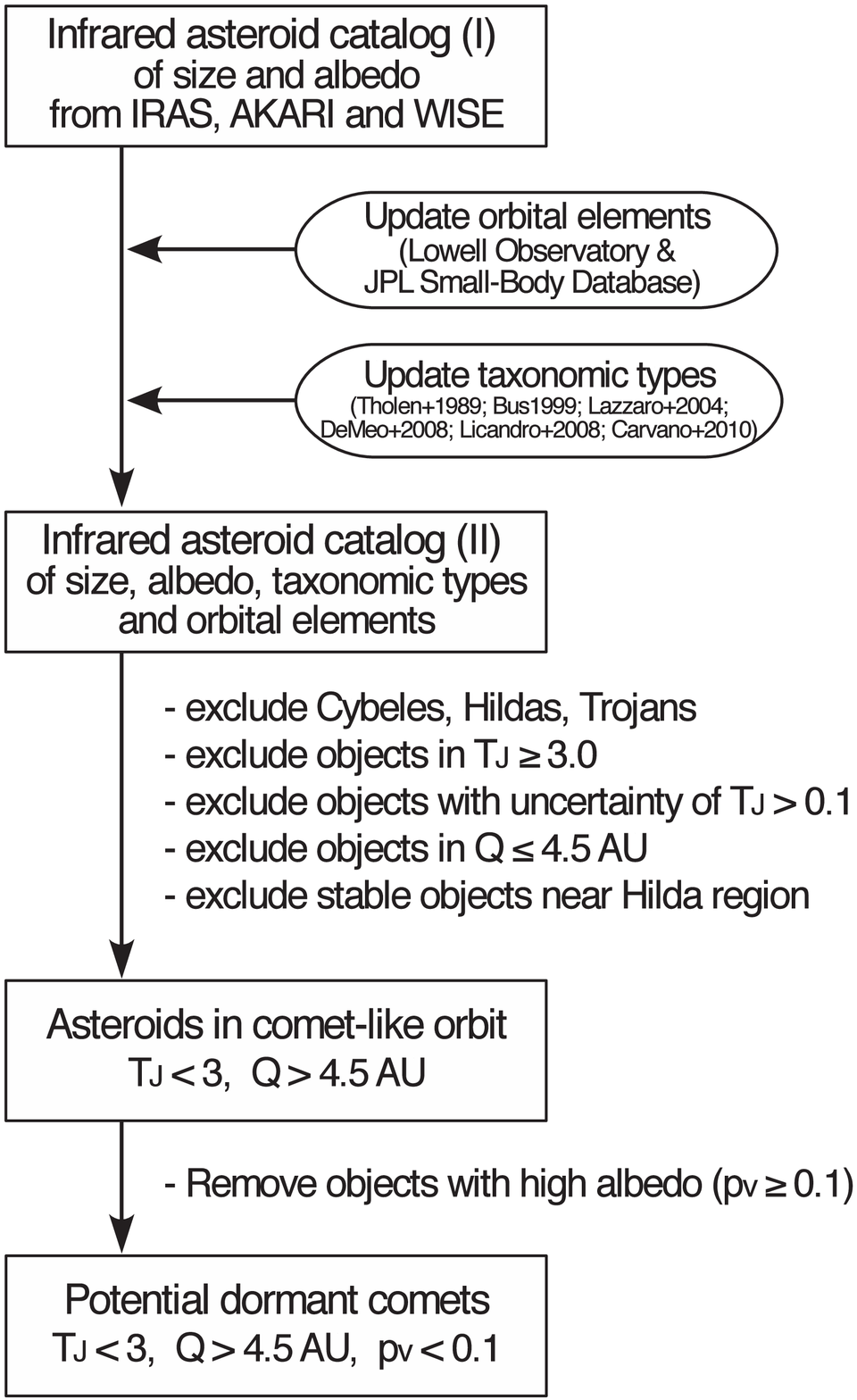} 
    \caption{Schematic diagram of the extraction flow. See Section 2.2.}
\label{fig:strategy}  
\end{center}   
\end{figure}

We summarize the extraction process in Figure \ref{fig:strategy}. The details are as follows.

\begin{enumerate}
%1111111
\item There are 138,285 asteroids whose albedos and sizes are  
given in the I--A--W catalog. We obtained the orbital
elements and spectral types of asteroids in the infrared catalogs. We added the orbital
elements of all 138,285 asteroids from the Lowell Observatory\footnote{ftp://ftp.lowell.edu/pub/elgb/astorb.dat} 
and the JPL Small-Body Database Browser.\footnote{http://ssd.jpl.nasa.gov/} In addition, we referred to spectral taxonomic types of asteroids,
if such information was available in \citet{Tholen1984}, \citet{Bus2002}, \citet{Lazzaro2004}, \citet{DeMeo2008},
\citet{Licandro2008}, and \citet{Carvano2010}.
After appending this ancillary  information, we obtained an infrared asteroid catalog of size
and albedo, together with spectral taxonomic types and orbital elements (see the top in Figure \ref{fig:strategy}).

%2222222
\item Second, we examined dynamical groups of asteroids according to their orbital elements. We followed the definition used in \citet{Zellner1985}, where they defined  dynamical groups of asteroids
based on their osculating orbital elements.
After classification of dynamical groups, we excluded asteroids in three dynamical groups --  Jupiter Trojans ($5.05 \le a \le 5.35$~AU), Hildas ($3.7 < a \le 4.2$~AU,
$e \le 0.3,$ and $i \le 20$\arcdeg), and Cybeles ($3.27 < a \le 3.70$~AU, $e \le 0.3,$ and $i \le 25$\arcdeg)
-- for extracting ACOs and PDCs, since these asteroids could be native objects trapped in
these regions soon after the formation of the Solar System and are unlikely to be dormant comets
recently captured in the current Solar System \citep{Marzari1998,Kortenkamp2001,Levison2008,Morbidelli2005}. 

%333333
\item We calculated the Tisserand parameter, $T_\mathrm{J}$. 
In general, a criterion of $T_\mathrm{J}<3$ has been applied for objects in comet-like
orbits whereas $T_\mathrm{J}>3$ has been applied for objects in asteroidal orbits,
although there are some exceptions for comets. Encke-type comets are visible active comets that
have the Tisserand parameter $T_\mathrm{J}>3.$ As of March 2014, there are only 39 comets
having $T_\mathrm{J}>3$ among  \textgreater600 short-period comets.
Main-belt comets have orbits indistinguishable from main-belt asteroids ($T_\mathrm{J}>3$) but show 
comet-like activities owing to sublimation of ice, impacts, and so on \citep{Jewitt2012,Hsieh2006}.
We excluded objects having $T_\mathrm{J} \ge 3$, and we do not consider  objects classified as
Encke-type comets and main-belt comets because of the difficulty in discriminating them from the majority
of asteroids.

%444444
\item Orbital uncertainty should be considered to extract ACOs from the catalog.
We noticed that there is a large discrepancy in the orbital elements of some asteroids between
Lowell Observatory and the JPL Small-Body Database Browser. In particular,
the \emph{WISE} mission discovered a number of new asteroids.
Some of them are newly discovered asteroids whose orbital elements are not determined
well because of inadequate orbital arcs and/or number of observations.
Since the Tisserand parameters calculated with poorly determined orbital elements could 
lead to erroneous results,  we computed the error of the Tisserand parameter, $\delta T_\mathrm{J}$,
and eliminated them from the list of ACOs unless the error was enough to distinguish ACOs from the majority
of asteroids. We thus calculated $\delta T_\mathrm{J}$ with the following equation:
\begin{equation}
\delta T_\mathrm{J}=\frac { \partial T_\mathrm{J} }{ \partial a } \delta a+\frac { \partial T_\mathrm{J} }{ \partial e } \delta e+\frac { \partial T_\mathrm{J} }{ \partial i } \delta i
,\end{equation}
where $\delta a$, $\delta e,$ and $\delta i$ are uncertainties of semimajor axis, eccentricity, and inclination, respectively, given by 
the JPL Small-Body Database Browser.
We set the threshold for the elimination of $\delta T_\mathrm{J}$ to be \textgreater0.1.
We excluded 349 ACO candidates with $T_\mathrm{J}< 3$ because of the large uncertainty in $\delta T_\mathrm{J}$.

%55555
\item In addition, we adopted a criterion for the aphelion of $Q>4.5$~AU following \citet{Fernandez2002}.
This is an effective condition to exclude some main-belt asteroids (e.g., asteroids with high-inclined orbits).
For comparison, we examined the orbital elements of active comets in the JPL Small-Body Database Search Engine\footnote{http://ssd.jpl.nasa.gov/sbdb\_query.cgi} 
and found that
all except 12 Encke-type comets and 10 main-belt comets have $Q> 4.5$~AU (as of March 2014).
Because we do not consider Encke-type comets and main-belt comets, we set the criterion $Q> 4.5$~AU.

%66666
\item Although we excluded the Hilda group above, 41 objects that marginally do not fall into the category of the Hilda group still remained.
Their osculating semimajor axes fall into the Hilda group (i.e., $3.7<a<4.2$~AU) but they have eccentricity and/or inclination
slightly larger than the Hilda asteroids.
As we discuss later, we excluded 38 objects among them and regarded only 3 objects as ACOs.

\end{enumerate}

\clearpage
\section{Results}
In this section, we examine the albedos of ACOs extracted based on the criteria in Section 2.2
and define the  PDCs that have comets-like orbits (i.e., $T_\mathrm{J}<3.0$ and $Q>4.5$~AU) 
and comet-like albedos ($p_\mathrm{v}< 0.1$).
Then, we study the physical characteristics of PDCs and the other ACOs.

\subsection{Albedo Properties of ACOs and Extraction of PDCs}

\begin{figure}
\begin{center} 
 \epsscale{0.55}
   \plotone{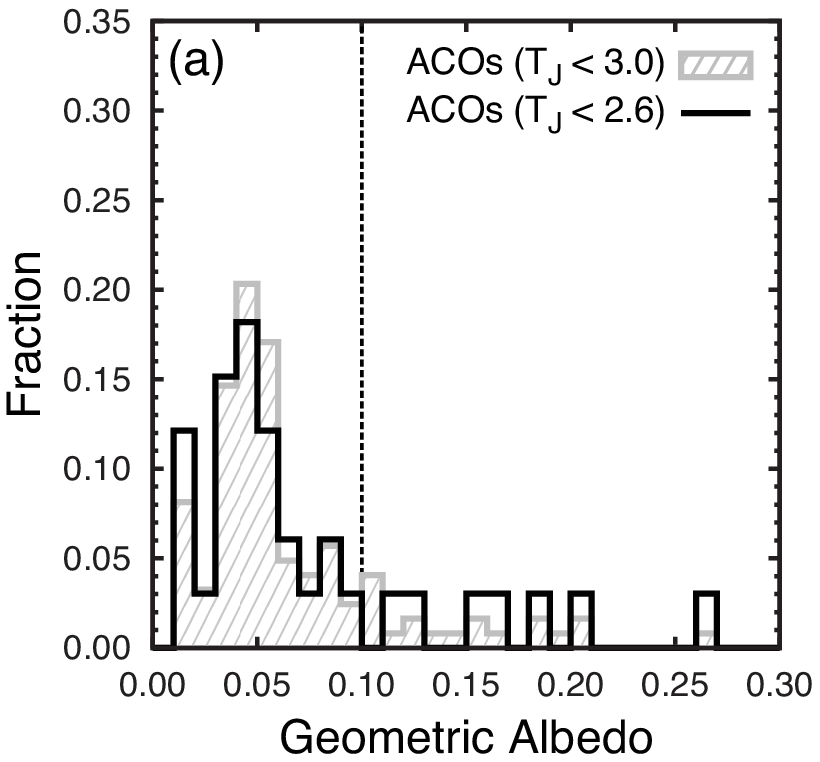} \plotone{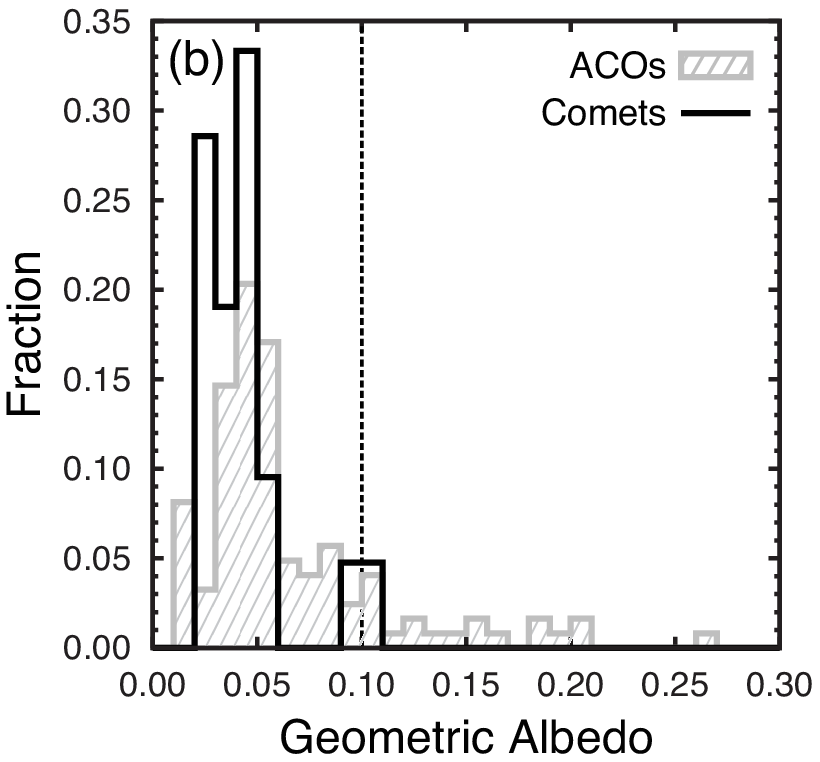}
  \caption{
Histogram of albedos of asteroids in comet-like orbits (ACOs).  
(a) Albedo distributions of ACOs with different $T_\mathrm{J}$ criteria, that is,
$T_\mathrm{J}<3$ and $T_\mathrm{J}<2.6$ following \citet{Fernandez2005}.
(b) Albedo distribution of ACOs and active comets. 
}\label{fig:albedo}   
\end{center}
\end{figure}

Figure \ref{fig:albedo}(a) shows the histogram of geometric albedos of ACOs. 
For comparison, we provide the albedo histogram  of ACOs with a
criterion for the Tisserand parameter, i.e., $T_\mathrm{J} < 2.6$,
 suggested by \citet{Fernandez2005} because it provides a more assured condition for extracting comet-like objects.
There is, however, no big differences between the two criterion of $T_\mathrm{J}$ in our sample of ACOs.
Since the appearance of histogram may depend on the choice of the bin width ($\Delta p_\mathrm{v}$),
we changed it from $\Delta p_\mathrm{v}$=0.005 to 0.015 but could not find any significant differences
in the appearances.
The histogram shows a prominent peak with a mode around 0.04--0.05.
Our sample of ACOs includes some objects with high albedos regardless of the severe $T_\mathrm{J}$ criterion.
Figure \ref{fig:albedo}(b) compares our sample of ACOs with that of active comets
based on the data in \citet{Lamy2004}, together with a few samples from \citet{Usui2011} and \citet{Bauer2013}.
Comets have a mean albedo of 0.046 with a standard deviation of 0.020
if we fit a Gaussian function.
Because no comets have albedo \textgreater\ 0.1, we placed the upper limit of comet-like albedo
at 0.1. Our upper bound of comet-like albedo (i.e., $p_\mathrm{v}=0.1$) is
slightly higher than that  in  \citet{Fernandez2001, Fernandez2005}, where they adopted the comet-like
albedo of $p_\mathrm{v}< 0.075$. The difference seems to be trivial, but we used the criterion of $p_\mathrm{v}= 0.1$
because a few comet nuclei with higher albedo ($p_\mathrm{v}>0.075$) were recognized 
($p_\mathrm{v}=0.096$ for 212P and $p_\mathrm{v}=0.101$ for C/2011 KP36) by \citet{Fernandez2005} and \citet{Bauer2013}.

The average albedo of PDCs is $p_\mathrm{v}=0.049 \pm 0.020$,
which is similar to those of the active comet nuclei mentioned above ($p_\mathrm{v}=0.046 \pm 0.020$). A peak in the
PDC distribution appears near 0.04--0.05, which is consistent with the peak in the comet
albedo distribution. These similarities suggest that we have extracted dormant comets from the
asteroid catalogs in an appropriate manner.

\subsection{Size Distribution of PDCs}

The size distribution of an ensemble of minor bodies gives us information helpful for  explaining
their source region and evolutionary history. The cumulative size distribution
has been applied in  previous research. It is approximated by the  mathematical form
\begin{equation}
N_{ S }(>D)\propto D^{ -q_S }
,\end{equation}
\noindent
where  
$N_{ S }(>D)$ is the cumulative number of bodies larger than diameter $D$, and $q_S$ denotes
the power index of the cumulative size distribution.

We examined the cumulative size distribution of PDCs in terms of two different origins:
Jupiter-family comets (JFCs),  considered to originate from the scattered disk of the Solar System
\citep{Levison1997}, and nearly-isotropic comets (NICs), originating from the Oort cloud
\citep{Weissman2002}. In some of the literature, NICs are further subdivided into long-period comets with orbital
period $P>200$~years and  Halley-type comets with $P<200$~years.
We separated our samples into two groups, one having $2 < T_\mathrm{J} < 3$
(PDCs in JFC-like orbits) and another having $T_\mathrm{J}<2$ (PDCs in NIC-like orbits). 

\begin{figure}
\begin{center} 
 \epsscale{0.55}
   \plotone{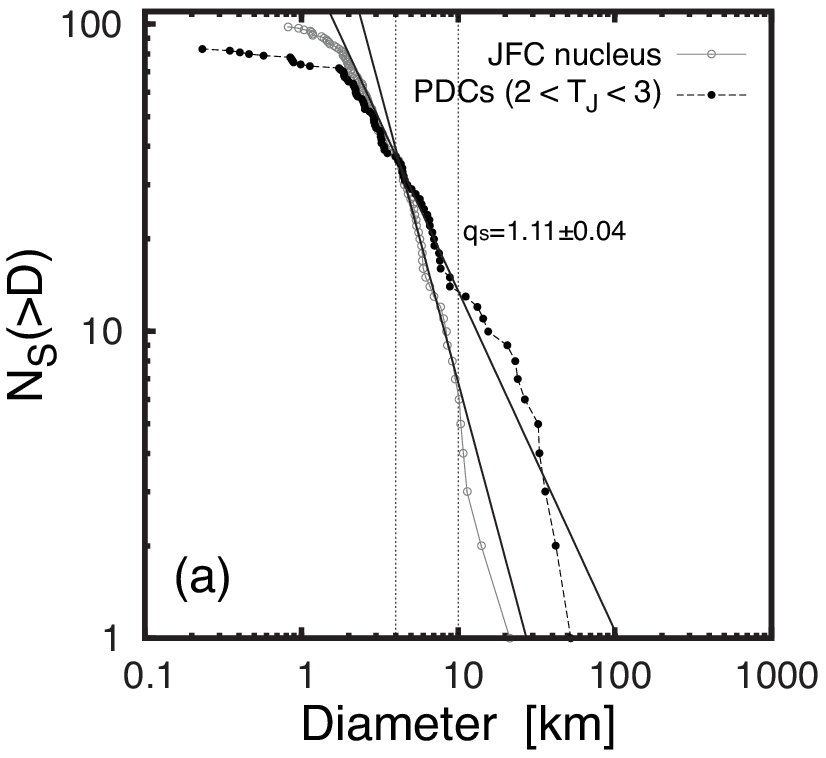} \plotone{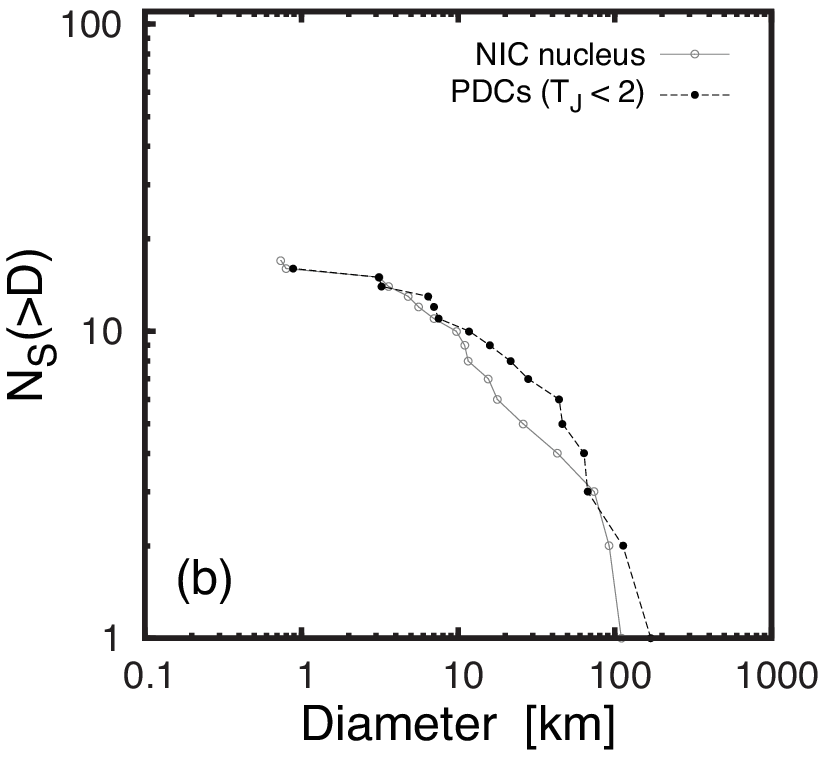}
  \caption{Cumulative size distribution of PDCs (filled circles) and ACOs (open circles).
  (a) Comparison between PDCs having $2 < T_\mathrm{J} < 3$  and active JFCs.
  (b) Comparison between PDCs having $T_\mathrm{J}<2$ and active NICs.}\label{fig:csd}     
\end{center}
\end{figure}

Figure \ref{fig:csd} shows the cumulative size distributions of PDCs in JFC-like orbits
(a) and PDCs in NIC-like orbits (b). We incorporate 11 PDCs from
\citet{Fernandez2001,Fernandez2005} in our list, so that the total number of PDCs
increased from 83 to 94. These size distributions are compared with those of active comets.
In Figure \ref{fig:csd} , we applied the sizes of active JFCs in \citet{Fernandez2013}
and active NICs in \citet{Lamy2004}. We found that PDCs have a shallower size distribution
than active comets, although the number of samples for NICs (only 17 PDCs in the NIC class) may
not be significant for a statistical discussion.
Careful comparison between PDCs in JFC-like orbits and active comets makes us aware
that the the size distributions show good agreement in the small-size range $2<D<4$~km,
that is, $q_S$ = 1.0 for PDCs and $q_S$ = 1.1 for JFC nucleus; however, they show a
significant discrepancy in  slope in the big-size range (4--10~km in diameter),
that is, $q_S$ = 1.11 $\pm$ 0.04 for PDCs and $q_S$ = 1.9 for active
JFCs \citep{Fernandez2013}.

\subsection{Spectral Types}
We have eight ACOs whose spectral taxonomic types are known.
There are one C-type object (7604 Kridsadaporn),
four D-type objects (944 Hidalgo, 3552 Don Quixote, 6144 Kondojiro, and 20898 Fountainhills), and
three X- or P-type objects (3688 Navajo, P/2006 HR30, and 2001 XP1).
The X-type \citep{Tholen1984} has subcategories depending on the albedo measurements: E- and M-types for high albedos and medium albedos and  P-type for the low albedos equivalent to comets (i.e., $p_\mathrm{v}<0.1$).  We classify two X-type ACOs as of P-type, according to their low albedos.
All eight objects have significantly low albedo $(p_\mathrm{v}<0.06$).
We examined the spectral slopes $S'$ of these eight objects, which express the percentage change
in the reflectance per 1000~\AA\ of wavelength difference \citep{Jewitt2002}. We computed the slope of
four objects (7604 Kridsadaporn, 944 Hidalgo, 3552 Don Quixote and 2001 XP1). Together with
the $S'$ values in \citet{DeMeo2008} and \citet{Licandro2008}, we obtained  $S'=5.7\pm4.8$,
which is consistent with the slopes of dormant comet candidates ($7.2\pm 2.0$\%) and comet nuclei
($8.3\pm2.8$\%)  \citep{Jewitt2002}.
Therefore, we conclude that these eight objects are most likely PDCs.
In fact, P/2006 HR30 and 3552 Don Quixote showed comet-like activities after their discovery as asteroids \citep{Hicks2007,Mommert2014A}.

\subsection{High-Albedo ACOs}
We identified a peculiar population of ACOs that have high albedos unlike comet
nuclei; this is an unexpected population. 
Because we excluded objects with $\delta T_\mathrm{J}>0.1$, they are not caused by inaccuracy in their orbital elements. The total divergence among I--A--W is 10\% in diameter
and 22\% in albedo
at the $1\sigma$ level \citep{Usui2014}, which is small enough to justify the existence of ACOs with high  albedo.
Therefore, it is likely that there are high-albedo asteroids
in comet-like orbits.

To understand the unexpected population, we studied the physical characteristics of these high-albedo ACOs.
Figure 4 shows a plot of diameter against perihelion distance for all ACOs (a) and ACOs with $T_\mathrm{J}<2.6$ (b).
Note that the paucity of small ACOs beyond $q\sim 2$~AU is an observational bias.
We found a pronounced tendency for high-albedo ACOs to concentrate in near-Earth space
(i.e., $q<1.3$~AU). In addition, they consist of small asteroids ($<$3 km). 
In Figure 4, there are four high-albedo ACOs at $q>1.3$~AU. Among them, three objects have albedo of $p_\mathrm{v}\sim 0.1$,
that is, 2009 QK35 ($p_\mathrm{v}=0.10 \pm 0.03$), 2010 MB86 ($p_\mathrm{v}=0.11 \pm 0.03$), and 2010 MK43 ($p_\mathrm{v}=0.10\pm 0.03$),
and one object, 2010 RM64 ($p_\mathrm{v}=0.16 \pm 0.05$),  marginally has high albedo. Thus, nearly all high-albedo ACOs consist of small asteroids
at $q<1.3$~AU. 
This trend cannot be explained by the observational bias. Because the result is obtained based on
the mid-infrared data, which, unlike optical observations, are less sensitive to albedo values, 
it provides reliable sets of asteroid albedo information. If there are big ACOs with high albedo beyond
$q=1.3$~AU, they would be detected easily. Although further dynamical study is essential to 
evaluate the population quantitatively, we propose that such ACOs with high albedos were injected
from the domain of $T_\mathrm{J}>3$ via the Yarkovsky effect, because small objects
with higher surface temperature are susceptible to the thermal drag force and gradually change their orbital
elements to be observed as ACOs in our list.

\clearpage
\section{Discussion}

\subsection{Comparison with Previous Research}

In this subsection, we compare our results with those from previous research on ACOs.
\citet{Fernandez2001, Fernandez2005} found that nearly all objects in their sample with $T_\mathrm{J}<2.6$ have comet-like
albedos ($p_\mathrm{v}<0.075$), implying that objects with $T_\mathrm{J}<2.6$ are most likely dormant comets.
They also found that  the transition region $2.6 < T_\mathrm{J} < 3$ comprises a mixture of high-albedo
and low-albedo objects, suggesting that the region is populated by a variety of sources.
A similar argument was made by \citet{Licandro2006, Licandro2008}, who insisted that a criterion
of $T_\mathrm{J} < 2.7$ provides the most reliable sets of PDCs based on their spectral survey of ACOs.
Following the work by \citet{Fernandez2001, Fernandez2005}, we made a plot of geometric albedo with respect to
$T_\mathrm{J}$ (Figure \ref{fig:TJ}). As suggested by \citet{Fernandez2001, Fernandez2005}, we confirm the similar trend that
most objects with small $T_\mathrm{J}$ have low albedo. However, our results  differ from those of \citet{Fernandez2001,Fernandez2005} 
and \citet{Licandro2006, Licandro2008} in that we find high-albedo objects in the region $T_\mathrm{J} < 2.6$. Because
\citet{Fernandez2001, Fernandez2005} detected small  objects ($<$3 km), it is not surprising that they potentially
detected ACOs with high albedo in the region of $T_\mathrm{J}<2.6$. It is not clear  why they did not detect
such ACOs with high albedo.
A possible explanation of such high-albedo objects is the inadequate measurements of optical magnitudes
in our samples. In fact, \citet{Fernandez2001, Fernandez2005} measured both optical and infrared magnitudes simultaneously
by themselves with ground-based telescopes, providing  reliable albedo data, whereas our infrared survey data relied on absolute magnitudes in which there is an intrinsic
difficulty in determining the absolute magnitudes and eventually the albedos \citep{Harris1997, Pravec2012}.

To refute the uncertain factor related to optical magnitudes, we
observed three objects: 2006 HY51 ($p_\mathrm{v} = 0.157 \pm 0.071$, $T_\mathrm{J} = 2.30$),
2006 CS ($p_\mathrm{v} = 0.037 \pm 0.021$, $T_\mathrm{J} = 2.44$), and
2010 NY1 ($p_\mathrm{v} = 0.037 \pm 0.008$, $T_\mathrm{J} = 2.66$).
We made optical observation of these ACOs with the Faint Object Camera and Spectrograph (FOCAS; \citealt{Kashikawa2002})
attached to the 8.2-m Subaru Telescope atop Mauna Kea (Hawaii, USA) on June 5, 2013 (UT). Flux calibration was performed using SA 104-430, SA 110-361, and MARK A2
listed in \citet{Landolt1992}. The observed magnitudes were converted into  absolute magnitudes by assuming a phase
slope of 0.035~mag~deg$^{-1}$ \citep{Lamy2004}. We obtained  absolute V magnitudes of 
{17.18 $\pm$ 0.30 (2006 HY51), 
16.76 $\pm$ 0.09 (2006 CS), and 17.07 $\pm$ 0.21 (2010 NY1)}, respectively.
These are consistent with the magnitude used by \citet{Mainzer2011a,Mainzer2012a},
who employed 17.20, 16.60, and 16.70, respectively.
Therefore, it is unlikely that optical magnitudes led
to poor albedo values, at least for these three ACOs.

With Subaru, we  derived the color of one object, 2006 CS ($p_\mathrm{v} = 0.037 \pm 0.021$), as $(\textrm{V}-\textrm{R}) = 0.341 \pm 0.126$ and
$(\textrm{B}-\textrm{V})=0.599 \pm 0.149$. These indices are less red than the average  comet nucleus but in the possible range of PDCs.
We could not derive the color of the other two objects because of the faintness at the time of our observation.
Instead, we found the spectrum of 2006 HY51 archived in SMASS.\footnote{http://smass.mit.edu/smass.html}
Although   2006 HY51's  taxonomic type is not identified,
it exhibits several key features typical of silicaceous (e.g., S-type) asteroids  having a red continuum at 1--1.5~\micron\ and shallow
absorption around 2~\micron, probably associated with the presence of pyroxene \citep{Gaffey1993}. Together with the
moderately high albedo ($p_\mathrm{v} = 0.157 \pm 0.071$ as determined from \textit{WISE})
and the spectrum, we may conclude that 2006 HY51 is not a dormant comet nucleus but is a high-albedo ACO.

\subsection{Asteroids in Comet-like Orbits near the Hilda Region}

In the I--A--W catalog, we found that there are a considerable number of objects
having $T_\mathrm{J} < 3$ but not categorized as ACOs by our criteria (see Section 2.2),
that is, 1764 in Jupiter Trojans, 432 in the Hilda region, and 104 in the Cybele region.
Theoretical studies suggested that the Trojans were captured during the early stages
of Jupiter's growth \citep{Marzari1998} or via the effects of nebular gas \citep{Kortenkamp2001}.
More recently, it has been suggested that they were captured during planetary migration, which occurred
about 500--600 million years after the Solar System's formation \citep{Levison2008,Morbidelli2005}.
In either case, it may be true that Jupiter Trojans were objects trapped in the early stage of the Solar
System's formation and not dormant comets captured recently.
The Hilda asteroids  are in another region  populated by interplanetary bodies in the 3:2 inner mean
motion resonance with Jupiter. Because of their stable orbits, the Hilda asteroids are also
considered to be objects formed in the early stage of the Solar System,
although there are about 10--20 comets that have been captured temporarily in the Hilda region \citep{Ohtsuka2008,Toth2006}.
The Hilda asteroids are defined as objects that have osculating orbital elements
$3.7 < a \le 4.2$~AU, $e \le 0.3,$ and $i \le 20$\arcdeg. As we described in Section 2.2,
there are a considerable number of objects (42) with $T_\mathrm{J} < 3.0$ that marginally
fall off the conditions of the Hilda group. 
Figure \ref{fig:hilda} shows the histograms of semimajor axis and eccentricity of ACOs,
including objects close to the Hilda asteroids. There is an unnatural prominent peak near the edge
of the Hilda region ($a \sim 4.0$~AU, $e \sim 0.3$).
We investigated the dynamical stability of 42 ACOs using the dynamical integration code Mercury 6 \citep{Chambers1999}. 
We thus integrated their orbits forward for 100,000 years (longer than the physical time scale of short-period comets).
%\textcolor{cyan}{Figure XX} shows examples of the orbital evolution.
We found that all  but three objects have stable orbits over the timescale of 100,000 years. Although there
are uncertainties in the dynamical simulation such as the value of the Yarkovsky force and the rocket force (for active comets),
we conservatively consider that  these three objects (2000 SU236, 2008 UM7, and 2009 SC298)  are ACOs and PDCs.

Let us consider how the Yarkovsky effect moves an asteroid into a comet-like orbit.  As shown in Figure 7(a), high-albedo
ACOs concentrate in a range of 2$<$a$<$3.5 AU, similar to main-belt asteroids and JFCs.  The Tisserand parameter is
a function of $a$, $e$, and $i$, while the Yarkovsky effect changes $a$. Due to the similarity in $a$ between high-albedo
ACOs and main-belt asteroids, we would conjecture that subsequent dynamical effects may change $e$ and $i$.
As widely known as a standard model for orbital evolution of near-Earth asteroids, the Yarkovsky effect could move small main-belt
asteroids' orbits until they are close to resonances with planets, and subsequently, these resonances can push them
in terrestrial planet crossing orbits \citep[see e.g.][]{Morbidelli2002}. Numerical simulations demonstrated that chaotic
resonances cause a significant increase in their $e$ and $i$ of test particles in the resonance regions \citep{Gladman1997}.
\citet{Bottke2002} suggested that some objects on $T_\mathrm{J}<$3 (or even $T_\mathrm{J}<$2) can result from
chaotic resonances. We reviewed the semimajor axes of ACOs in our list. Figure 8 shows $a$-$e$ plot of ACOs together
with major resonances. Although there are a couple of ACOs close to resonances, their semimajor
axes are not related to these major resonances. Therefore, it may be reasonable to think that encounters with terrestrial planets
as well as chaotic resonances with massive planets can drift main-belt asteroids into comet-like orbits.

\subsection{Discussion of the Size Distribution of PDCs}
Over the past decade,  a number of attempts have been made to derive the size distributions of active comets
\citep{Lamy2004,Meech2004,Snodgrass2011,Fernandez2013} and dormant comets \citep{Alvarez2006,Whitman2006}.
These research results exhibit a wide range of  the power index of the cumulative size distribution, that is, $q_S=1.5$--2.7 for active comets and $q_S=1.5$--2.6 for 
dormant comets. Most of this research (except that of \cite{Fernandez2013}) was conducted by using optical magnitudes of these objects.
Note that the observed quantity (i.e., magnitude) is proportional to the product of the albedo and the square of the size,
and in these research studies albedo values were assumed. In addition, the phase-angle dependence of magnitudes is also presumed in these studies.
Since the data taken in the mid-infrared wavelength provide the size of the objects without any albedo assumption, they yield a reliable data set of the size
distribution of active comets and dormant comets.

We therefore compare the power exponents of active JFCs,  $q_S \sim1.9$ \citep{Fernandez2013}, with
that of PDCs, $q_S=1.11\pm 0.04$ (this work). Because the difference is significantly larger than those of the errors of measurements,
this indicates that a physical mechanism is responsible for creating the difference in the course of cometary evolution.
The moderate slope of the size distribution for PDCs ($q_S=1.11 \pm 0.04$) implies that big objects are more abundant
in the list of dormant objects.  
The splitting of comets, a phenomena that has been occasionally observed, can be a possible mechanism leading to the change in
the power exponents. A good example event was observed at 73P/Schwassmann-Wachamann 3, in which the comet nucleus was ground 
into 10--100~m bodies. The power exponents of the cumulative size distribution of the fragments was  $\sim\negthickspace-2.3$ \citep{Ishiguro2009,Fuse2007}.
If the splitting is a principal mechanism in determining the size distribution of comet nuclei, the size distribution of
PDCs might be steeper than the value we derived for PDCs.
Another explanation is that big comet nuclei may develop inert surface dust layers more effectively than small ones
because dust particles with small ejection velocity cannot  escape from the big nuclei against the gravitational force
and eventually form surface dust layers that insulates subsurface ice against solar heating and/or choke off ice sublimation.
A similar argument was made by \citet{Tancredi2006}, who mentioned the possible choking mechanism on the large comet
nuclei of 49P/Arend--Regaux and 28P/Neujmin 1.

We also should consider the observational bias. Big objects are favorably detected at distant locations. Figure 7(a) shows
the diameter of ACOs as a function of the semimajor axis $a$. In the figure, it is clear that only big ACOs were detected beyond
$a \sim 3.5$~AU, most likely because of the observation selection bias. We  plotted the size distribution of PDCs within $a<3.5$~AU
to evaluate the effect. Figure 7(b) shows a comparison of the size distribution between all PDCs and PDCs within $a<3.5$~AU.
Although there is a large discrepancy in the largest size probably because of the insignificant number of samples, the slopes are 
consistent with one another in the size range of 2--4~km. In addition, we adopted a criterion to eliminate the observational bias
in \citet{Alvarez2006}, and obtained $q_S$=1.27. Although the power exponent is sensitive to the criteria we selected,
we can safely say $q_S$ is around 1.1 (neither $<$0.8 nor $>$1.3).
Therefore, we may rule out the possibility of  observational selection bias.
In summary, rapid growth of a dust mantle on the big comet nuclei favors the obtained size exponent of PDCs.

\clearpage
\section{Summary}

In this paper, we address the question of the existence of dormant comets in the list of 
asteroids. Motivated by the recent developments of infrared asteroid catalogs (I--A--W)
taken with three infrared surveyors \citep{Usui2014}, we identified ACOs with the criteria we 
contrived, i.e., $T_\mathrm{J} < 3.0$ and $Q > 4.5$~AU.
The major findings of our research are as follows:

\begin{enumerate}
\item There are 123 ACOs in the I--A--W catalog after rejection of objects with large orbital uncertainties.
\item The majority ($\sim \negthickspace80$\%) of ACOs have comet-like albedo (i.e., $p_\mathrm{v}<0.1$).
\item Low-albedo ACOs (referred to as PDCs) have an albedo distribution similar to that of active comets,
        that is, $p_\mathrm{v}=0.049 \pm 0.020$. 
        They have a shallower size distribution than that of active comets (i.e. $q_S \sim 1.1$).
\item Nearly all high-albedo ACOs consist of small bodies distributed in near-Earth orbit.
\end{enumerate}

In particular, we  stress again the significance of high-albedo ACOs. As we discussed through our
ground-based observation with the Subaru Telescope, high-albedo ACOs, which may have composition
similar to silicaceous asteroids, definitively exist in the I--A--W database. Considering the very low $T_\mathrm{J}$ as
well as the small size and perihelion distance, we would suggest that such high-albedo ACOs have been injected
via nongravitational forces, most likely the Yarkovsky effect.

\acknowledgments

\vspace{1cm}
{\bf Acknowledgments}\\
This work was supported by the National Research Foundation of Korea
(NRF)  funded by the South Korean government (MEST) (Grant No. 2012R1A4A1028713).
This study is based on observations with \textit{AKARI}, a JAXA project with
the participation of ESA.
This work also makes use of data products from
the \emph{Wide-field Infrared Survey Explorer}, which is a joint
project of the University of California, Los Angeles, and the Jet
Propulsion Laboratory/California Institute of Technology, funded
by the National Aeronautics and Space Administration.
Optical data were partially collected at the Subaru Telescope, which is operated
by the National Astronomical Observatory of Japan.
We thank  T. Kasuga and R. Brasser for their valuable comments and T. Hattori for supporting observations at Subaru.
We also thank anonymous referee for valuable comments. MI would like to express his gratitude to D. Jewitt for a support
during his stay at UCLA.
Part of the data utilized in this publication were obtained from SMASS,
which were made available by the MIT-UH-IRTF Joint Campaign for NEO Spectral Reconnaissance.
%The IRTF is operated by the University of Hawaii under Cooperative Agreement no. NCC 5-538 with the National Aeronautics and Space Administration, Office of Space Science, Planetary Astronomy Program. The MIT component of this work is supported by NASA grant 09-NEOO009-0001, and by the National Science Foundation under Grants Nos. 0506716 and 0907766.

\clearpage
%%%%%%%%%%%%%%%%%
%    FIGURE 1   %
%%%%%%%%%%%%%%%%%
\clearpage

\clearpage
%%%%%%%%%%%%%%%%%
%    FIGURE 2   %
%%%%%%%%%%%%%%%%%
\clearpage

\clearpage
%%%%%%%%%%%%%%%%%
%    FIGURE 3   %
%%%%%%%%%%%%%%%%%
\clearpage

\clearpage
%%%%%%%%%%%%%%%%%
%    FIGURE 4   %
%%%%%%%%%%%%%%%%%

\begin{figure}
\begin{center} 
 \epsscale{0.55}
   \plotone{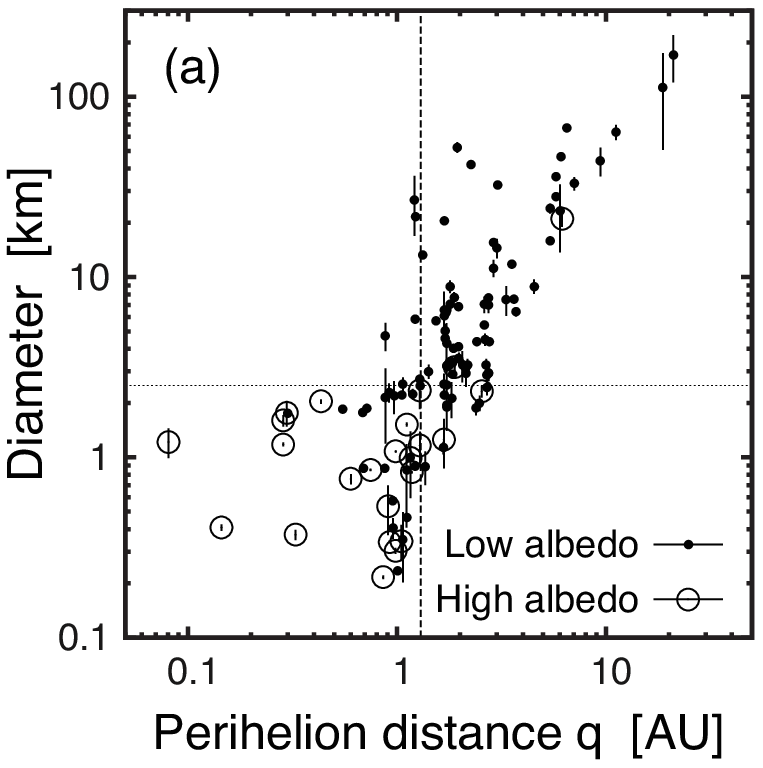}  \plotone{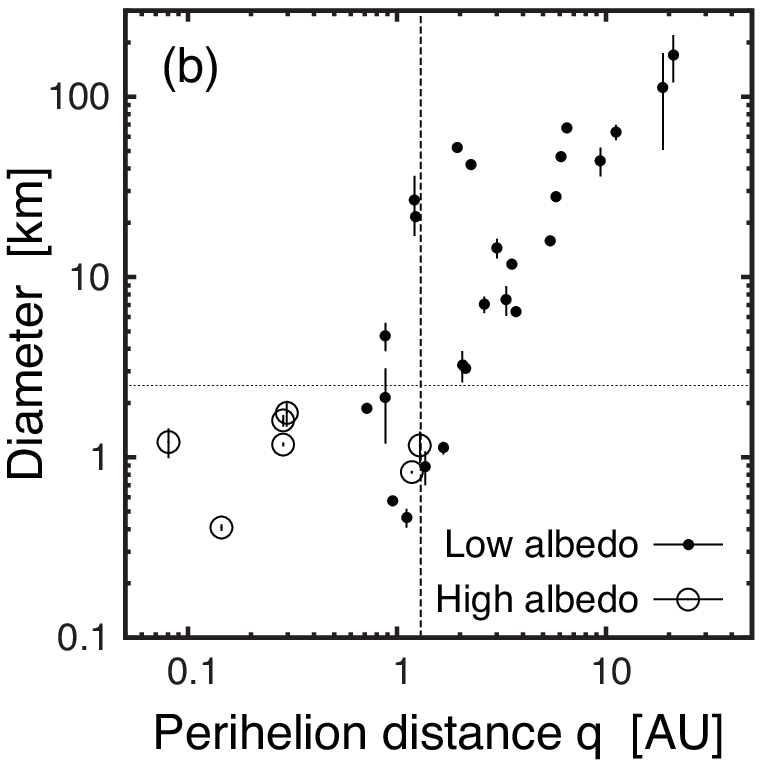} 
  \caption{
Plot of the perihelion distance $q$ vs. the diameter of all ACOs (a) and ACOs with $T_\mathrm{J}<2.6$ (b).
The open circles denote the high-albedo objects ($p_\mathrm{v} \ge 0.1$);  the
filled circles denote the low-albedo objects ($p_\mathrm{v} <0.1$).  We draw two lines for the 
perihelion distance at 1.3~AU (vertical line) and the diameter at 2.5~km (horizontal line) to discriminate small near-Earth objects (lower left).}
\label{fig:qd}     
\end{center}
\end{figure}
\clearpage

\clearpage
%%%%%%%%%%%%%%%%%
%    FIGURE 5   %
%%%%%%%%%%%%%%%%%
\begin{figure}
    \centering
 \epsscale{0.55}
   \plotone{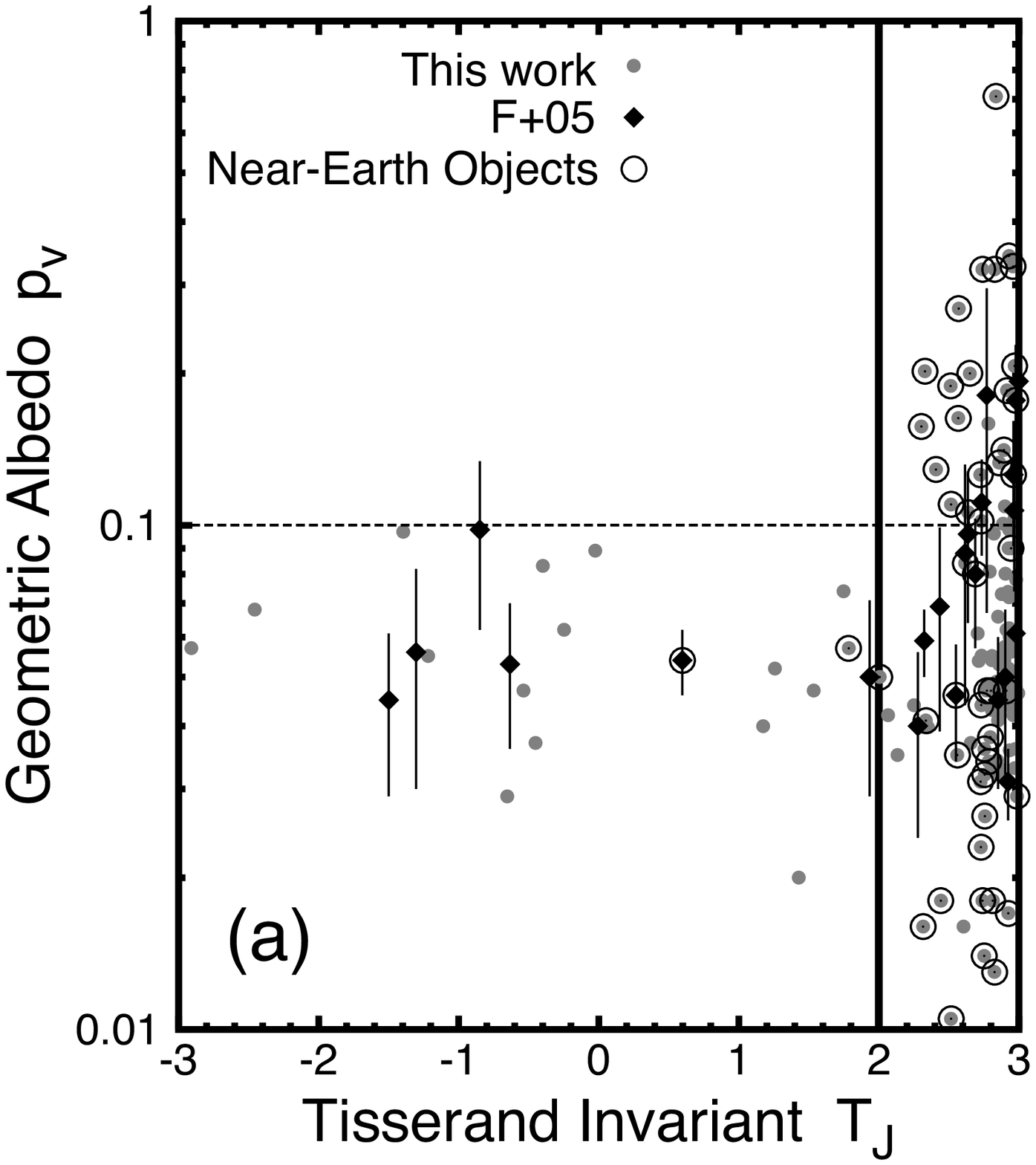} \plotone{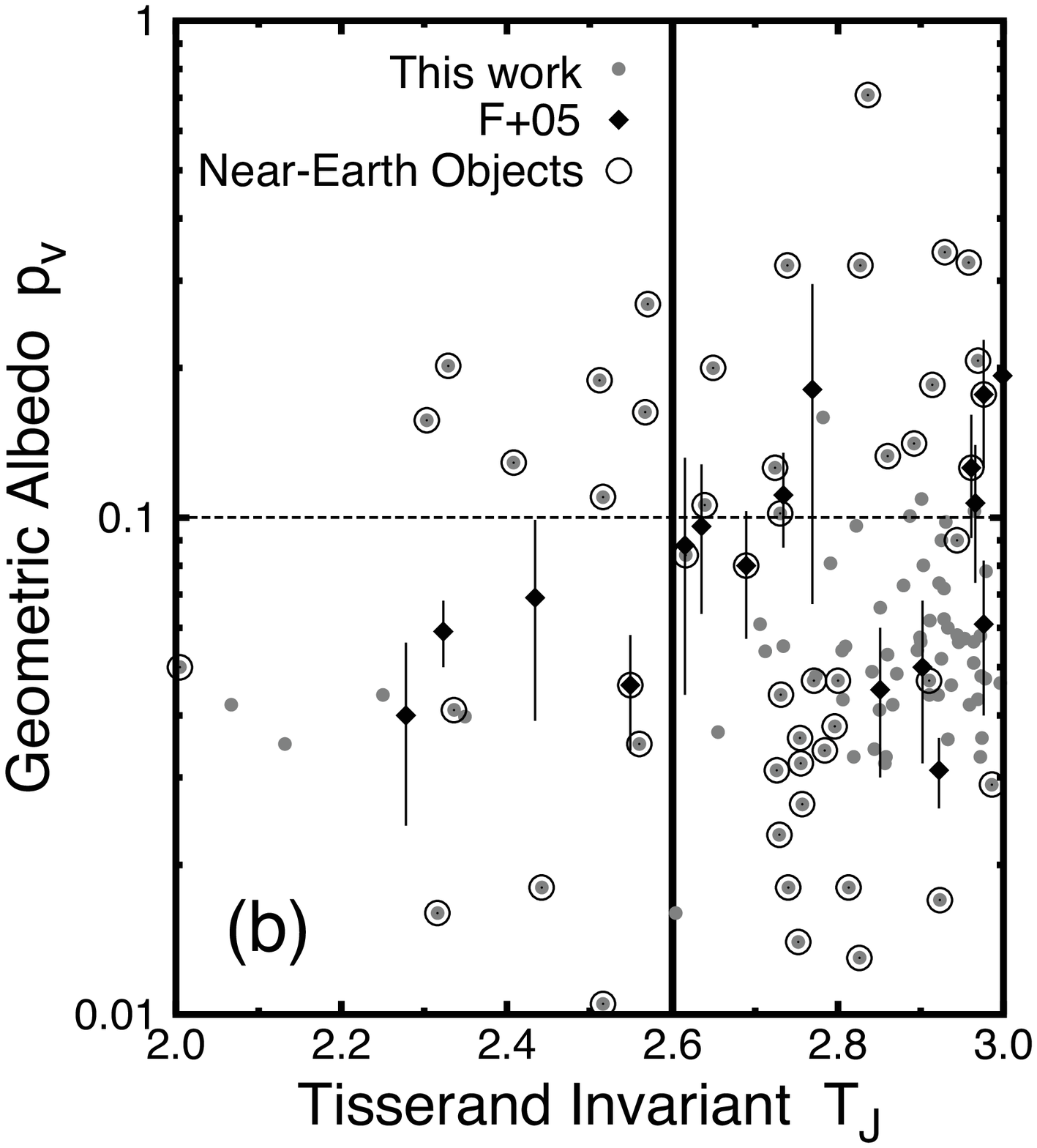}
  \caption{
Plots of the Tisserand parameter $T_\mathrm{J}$ vs. geometric albedos in the range  of 
$-3<T_\mathrm{J}<3$ (top) and $2<T_\mathrm{J}<3$ (bottom).
For comparison, we show the result in \citet{Fernandez2005}, which is indicated
by `F+05' in the figures. Data points of near-Earth objects are enclosed by circles.
The horizontal line corresponds to $p_\mathrm{v} =0.1$.
}\label{fig:TJ}     
\end{figure}
\clearpage

\clearpage
%%%%%%%%%%%%%%%%%
%    FIGURE 6   %
%%%%%%%%%%%%%%%%%
\begin{figure}
    \centering
 \epsscale{0.55}
   \plotone{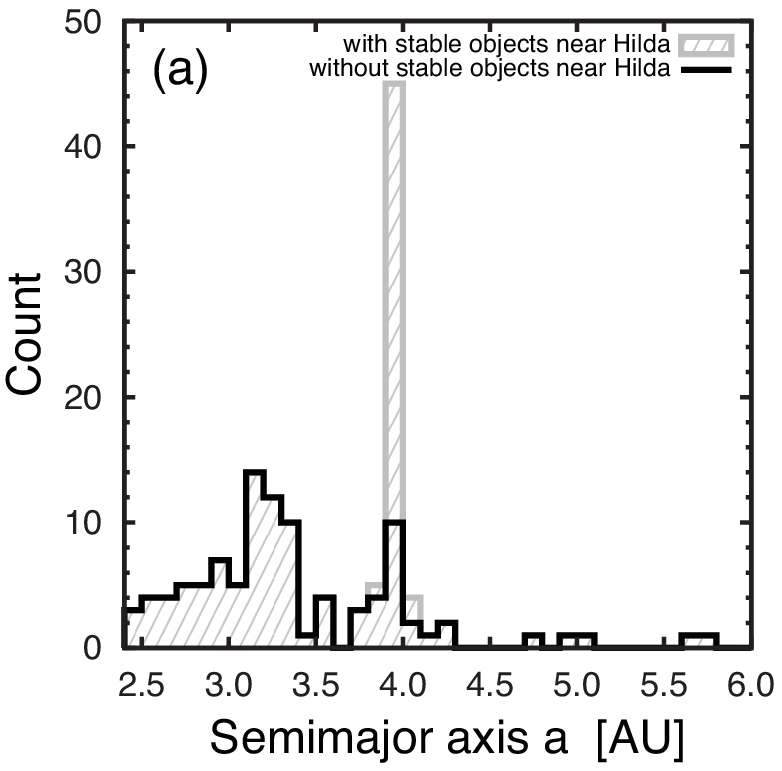} \plotone{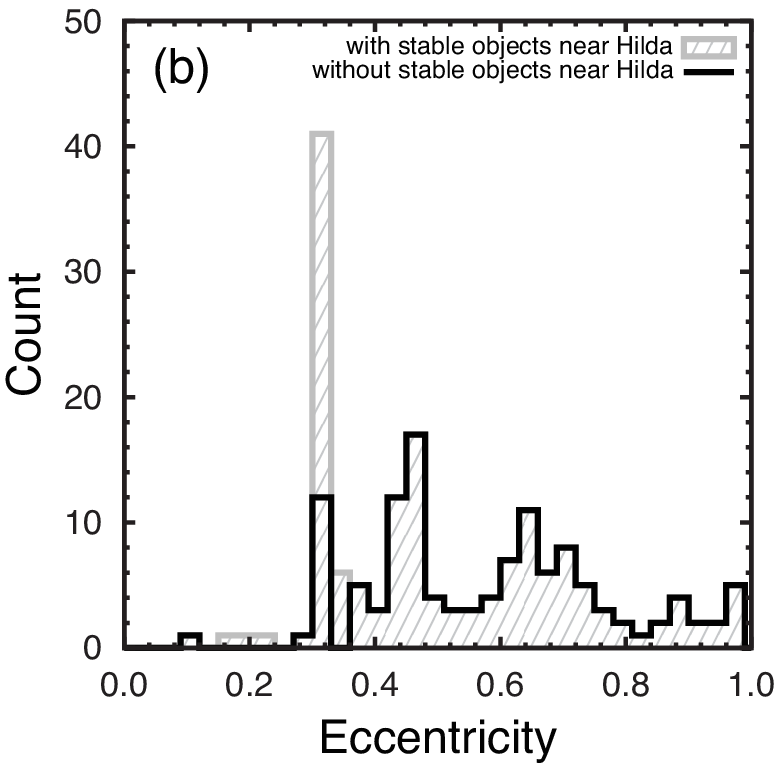}
  \caption{
Histograms of semimajor axis and eccentricity of ACOs
including the Hilda-like objects. There is a unnatural prominent peak near the Hilda regions (see Section 4.2).}\label{fig:hilda}     
\end{figure}
\clearpage

\clearpage
%%%%%%%%%%%%%%%%%
%    FIGURE 7   %
%%%%%%%%%%%%%%%%%
\begin{figure}
\begin{center} 
    \epsscale{0.60}
   \plotone{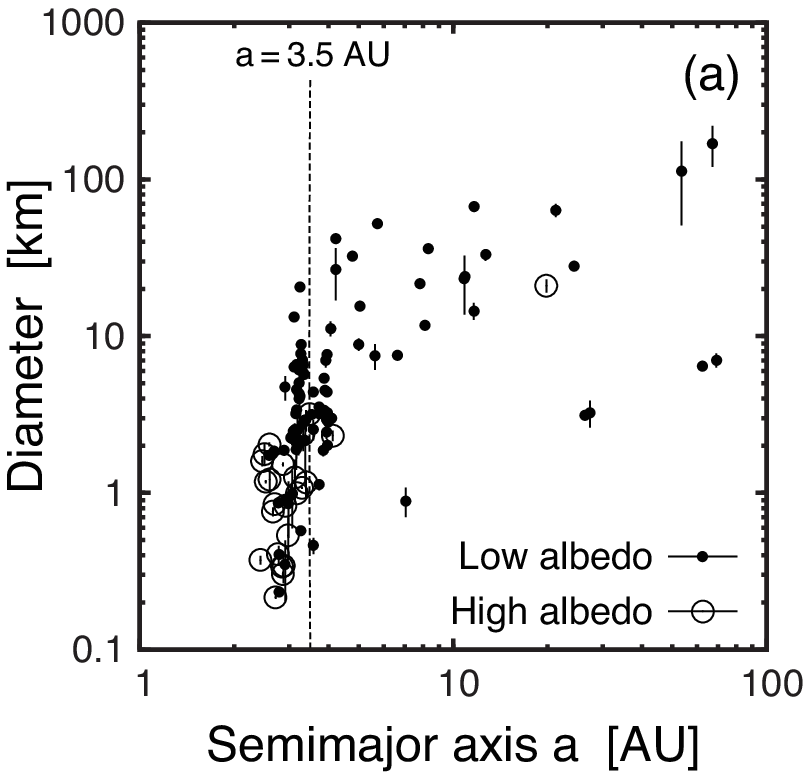}  \plotone{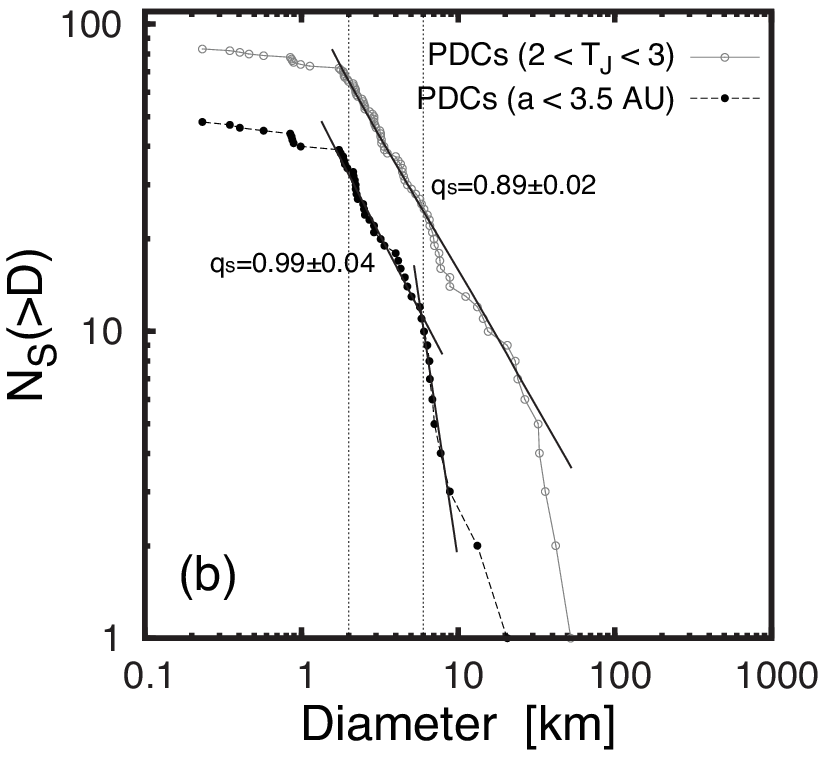} 
    \caption{(a) Plot of the semimajor axis $a$ vs. the diameter of asteroids in comet-like orbits.
Open circles denote the high-albedo objects ($p_\mathrm{v} \ge 0.1$);  
filled circles denote the low-albedo objects ($p_\mathrm{v} <0.1$). Small ($\lesssim$3 km) objects were detected within $\lesssim$3.5 AU.
(b) Cumulative size distribution of PDCs. 
Open circles stand for all PDCs;  filled circles stand for PDCs with $a<3.5$~AU. }
\label{fig:CSD2}  
\end{center}   
\end{figure}
\clearpage

\clearpage
%%%%%%%%%%%%%%%%%
%    FIGURE 8   %
%%%%%%%%%%%%%%%%%
\begin{figure}
\begin{center} 
    \epsscale{0.60}
   \plotone{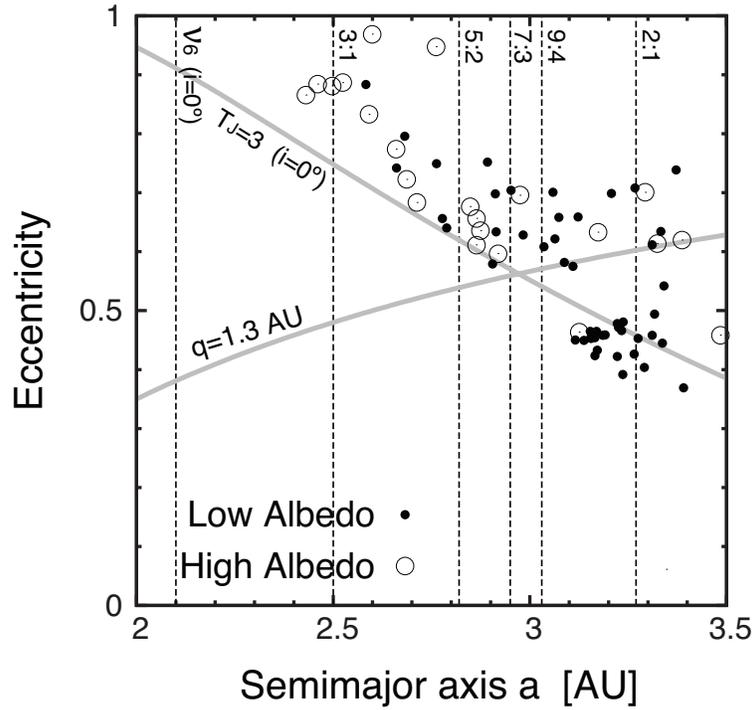} 
    \caption{(a) Plot of the semimajor axis $a$ vs. the eccentricity $e$ of asteroids in comet-like orbits.
Open circles denote the high-albedo objects ($p_\mathrm{v} \ge 0.1$);  
filled circles denote the low-albedo objects ($p_\mathrm{v} <0.1$). 
A line of $q$=1.3 AU and $T_\mathrm{J}=3.0$ ($i$=0\arcdeg\ is assumed) are drawn to 
clarify near-Earth objects and JFC-like objects. Vertical lines correspond to major resonances with planets.}
\label{fig:a-e-plot}  
\end{center}   
\end{figure}
\clearpage

%%%%%%%%%%%%%%%%%
%    TABLE 1    %
%%%%%%%%%%%%%%%%%

\begin{table}
\begin{center}
\footnotesize
\caption{Astroids in Comet--like Orbits in Infrared Asteroidal Survey Catalogs \label{tbl-aco}}
\begin{tabular}{lccccccccc}
\\
\tableline\tableline
Number  &  Name  &  Desig  &  $a$  &  $e$  &  $i$  &  $T_\mathrm{J}$  &  $D$  &  $p_\mathrm{v}$  &  Source  \\
%No.  &  Name  &  Desig.  &  $a$  &  $e$  &  $i$  &  $T_\mathrm{J}$  &  $D$  &  $e_\mathrm{D}$  &  $p_\mathrm{v}$  &  $e_pV$  &  Source  \\
      &             &               &  [AU]  &           & [deg]  &                             &  [km]  &          & \\
%      &             &               &  [AU]  &           & [deg]  &                             &  [km]  &         [km]                &           &    & \\
\tableline
944 & Hidalgo & 1920 HZ & 5.737 & 0.662 & 42.54 & 2.067 & 52.450 $\pm$ 3.600 & 0.042 $\pm$ 0.007 & \emph{ AKARI} \\
1922 & Zulu & 1949 HC & 3.237 & 0.481 & 35.42 & 2.734 & 20.561 $\pm$ 0.321 & 0.055 $\pm$ 0.006 & \emph{ WISE} \\ 
3688 & Navajo & 1981 FD & 3.222 & 0.478 & 2.56 & 2.996 & 6.086 $\pm$ 0.051 & 0.047 $\pm$ 0.012 & \emph{ WISE} \\ 
3552 & Don Quixote & 1983 SA & 4.222 & 0.713 & 30.96 & 2.316 & 26.656 $\pm$ 9.734 & 0.016 $\pm$ 0.009 & \emph{ WISE} \\ 
5370 & Taranis & 1986 RA & 3.333 & 0.634 & 19.09 & 2.731 & 5.821 $\pm$ 0.300 & 0.044 $\pm$ 0.009 & \emph{ WISE} \\ 
\end{tabular}
\tablecomments{This table is available in its entirety in a machine--readable form in the online journal. A portion is shown here for guidance regarding its form and content.}
\end{center}
\end{table}
\clearpage

\end{document}